\documentclass{article}

\PassOptionsToPackage{numbers, compress}{natbib}

\usepackage[preprint]{nips_2018}

\usepackage[utf8]{inputenc} 
\usepackage[T1]{fontenc}    
\usepackage{hyperref}       
\usepackage{url}            
\usepackage{booktabs}       
\usepackage{amsfonts}       
\usepackage{nicefrac}       
\usepackage{microtype}      
\usepackage{subcaption}
\usepackage{graphicx}
\usepackage[numbers]{natbib}

\title{Identifying viruses from metagenomic data by deep learning}

\author{
	Jie Ren$^*$ \\
    University of Southern California \\
	Los Angeles, CA, USA 
	\And
	Kai Song\thanks{co-first author} \\
	Chinese Academy of Sciences \\
	Qingdao, China 
	\And
	Chao Deng \\
	University of Southern California \\
	Los Angeles, CA, USA 
	\And
	Nathan A. Ahlgren \\
	Clark University \\
	Worcester, MA, USA  
	\And
	Jed A. Fuhrman \\
	University of Southern California, \\
	Los Angeles, CA, USA 
	\And
	Yi Li \\
	University of California, Irvine \\
	Irvine, CA, USC  
	\And
	Xiaohui Xie \\
	University of California, Irvine \\
    Irvine, CA. USA
	\And
	Fengzhu Sun\thanks{corresponding author, fsun@usc.edu} \\
	University of Southern California \\
	Los Angeles, CA, USA \\
	Fudan University \\ 
	Shanghai, China
}

\begin{document}

\maketitle

\begin{abstract}
The recent development of metagenomic sequencing makes it possible to sequence microbial genomes including viruses in an environmental sample.
Identifying viral sequences from metagenomic data is critical for downstream virus analyses.
The existing reference-based and gene homology-based methods are not efficient in identifying unknown viruses or short viral sequences. 
Here we have developed a reference-free and alignment-free machine learning method, DeepVirFinder, for predicting viral sequences in metagenomic data using deep learning techniques.
DeepVirFinder was trained based on a large number of viral sequences discovered before May 2015.
Evaluated on the sequences after that date, DeepVirFinder outperformed the state-of-the-art method VirFinder at all contig lengths. 
Enlarging the training data by adding millions of purified viral sequences from environmental metavirome samples significantly improves the accuracy for predicting under-represented viruses.
Applying DeepVirFinder to real human gut metagenomic samples from patients with colorectal carcinoma (CRC) identified 51,138 viral sequences belonging to 175 bins. Ten bins were associated with the cancer status, indicating their potential use for non-invasive diagnosis of CRC. 
In summary, DeepVirFinder greatly improved the precision and recall rates of viral identification, and it will significantly accelerate the discovery rate of viruses.

\end{abstract}

\section{Introduction}

Studies of the effect of viruses on their host communities as well as on public health are in their very earliest stages. It has only recently become possible to identify viruses in large metagenomic datasets using next-generation sequencing (NGS) technologies and computational advances. Unlike traditional methods of isolating viruses through laboratory cultures, metagenomic sequencing effectively sequence all genetic materials in a microbial community regardless the cultivability, exhibiting true viral diversity and accelerating the rate of virus discovery. A few studies used metagenomics sequencing technology to study viruses in human gut, and revealed the important association between viruses and human diseases, such as inflammatory bowel disease (IBD) \cite{norman2015disease}, Severe Acute Malnutrition (SAM) \cite{reyes2015gut}, and type II diabetes \cite{ma2018human}.


Identifying viruses from metagenomic samples is the first crucial step for all the downstream viral analyses. Different computational methods have been developed to tackle the virus identification problem. The tools, Metavir \cite{roux2011metavir} and ViromeScan \cite{rampelli2016viromescan}, classify viral metagenomics reads based on sequence alignment and homology searches against known viral reference genomes. Kraken \cite{wood2014kraken}, Centrifuge \cite{kim2016centrifuge}, and MetaPhlAn \cite{truong2015metaphlan2}, though designed mainly for mapping bacteria reads to known genome database, also include viral reference genomes and can rapidly map reads to known viruses. However, known viruses in the current database are markedly biased towards certain types whose hosts are cultivable in lab. The current virus references are far from being representative of the whole viral diversity. It is estimated that only about 15\% of viruses in human gut have similarity to known viruses in the database.

VIROME \cite{wommack2012virome}, DIAMOND \cite{buchfink2015fast}, VirSorter \cite{roux2015virsorter}, and IMG/VR \cite{paez2017nontargeted,paez2016uncovering} identify viral sequences by comparing genes in the sequence with viral gene database, and see if there is enough evidence showing the query sequence carries viral proteins. The advantage of comparing sequences at amino acid level instead of nucleotide level is that, amino acids tolerate more mutations so that the prediction are more stable, since viruses have high mutation rate. In addition, VIROME, VirSorter. and IMG/VR enlarged the viral protein database by adding viral proteins from metavirome datasets to detect unknown viruses. The protein based methods normally require query sequences contains complete genes, and VirSorter requires at least three genes to make predictions to achieve high accuracy. However, since metagenomics reads and metagenomically assembled sequences (called contigs) are mainly about hundreds of base pairs, most sequences do not have complete genes. Therefore, the gene homology based methods exclude many short contigs. 

Recently we developed a new program, VirFinder, to identify viral sequences using a machine learning method based on features of k-mer frequencies \cite{ren2017virfinder}. VirFinder automatically learned k-mer patterns enriched/depleted in viruses and built a powerful classifier to predict viral sequences. Since VirFinder characterizes sequences using k-mer frequencies, it can predict contigs from both coding and non-coding regions. At the same time, it makes full use of information from both coding and non-coding regions for prediction. Moreover, VirFinder does not depend on gene finding and homology searches, so it can predict short contigs containing only partial genes or not enough numbers of genes. The success of VirFinder demonstrates that the machine learning based methods are more powerful than traditional alignment-based methods, and there are more potentials to be explored.

Deep learning is one type of machine learning algorithms based on artificial neuron networks. It is inspired from the biological neural networks that transform the input to output by processing signals through a cascade of multiple layers of neurons. Thanks to exponentially increasing computing power and big data, deep learning wins great success in computer vision, speech recognition, and natural language procession. Recently, a few studies developed deep learning methods to solve different problems in computational biology, such as predicting the sequence specificity of protein binding using deep learning \cite{alipanahi2015predicting, zeng2016convolutional, quang2017factornet, wang2018define}, predicting the effect of noncoding variants based on different model structures \cite{zhou2015predicting, quang2016danq}, predicting chromatin accessibility \cite{kelley2016basset}, calling genetic variants from millions of short reads in NGS samples \cite{poplin2017creating}, predicting methylation quantitative trait loci \cite{zeng2017predicting}, identifying evolutionarily conserved sequences \cite{li2017understanding}, and identification of enhancer and promoter \cite{li2016genome} and their interactions \cite{singh2016predicting}. See \citep{yue2018deep} for detailed reviews of deep learning applications in genomic research. Those methods have shown remarkable improvements over the traditional machine learning-based and statistical inferences-based models. 

Those methods commonly use convolutional neural networks (CNN or ConvNet), a class of deep neuron networks, to automatically learns genomic signatures from the input sequences, and simultaneously build a predictive model based on the learned genomic signatures. 
Convolutional networks first recognize different patterns of a fixed length in the input sequences, and summerize the pattern intensities along the genome using a summary statistic. The patterns in the context of genomic sequence, are similar to the biological motifs. A motif can be represented using a position weight matrix which specifies the distribution of having 4 letters A, C, G, and T at each position. When the distribution at each position is a constant, i.e. one letter has probability 1 and the others have 0 probability, the motif degenerates to be a k-mer. Thus, the ConvNet is a natural generalization of k-mer based model. Using the more flexible deep convolutional neural networks can provide model with higher prediction accuracy than k-mers based models.

In this study, we developed a new method DeepVirFinder, to the best of our knowledge, the first deep learning based program for identifying viral sequences from metagenomic data. Trained using viral reference genomes, DeepVirFinder efficiently learned a convolutional neural network that accurately identified viral sequences, and it outperforms the previous state-of-the-art method VirFinder at all contig lengths.
To further elevate the representation of minor viral groups, we enlarged the training data by adding millions of viral contigs in metavirome datasets. 
The model with enlarged dataset had significantly higher prediction accuracy for the under-represented viral groups. 
As a case study, we applied DeepVirFinder to identify viruses in the human gut metagenomic samples from patients with colorectal cancer. Ten viruses associated with the cancer status were discovered, suggesting the potential of using viruses as a non-invasive diagnostic testing for colerectal cancer.

\section{Materials and Methods}
\subsection{Viruses and prokaryotic genomes used for training, validation and testing}

We collected 2,314 reference genomes of viruses infecting prokaryotes (bacteria and archaea) from NCBI (\href{}{https://www.ncbi.nlm.nih.gov/genome/browse}). The dataset was partitioned into three parts based on the dates when the genomes were discovered. We used the genomes discovered before January 2014 for training, those between January 2014 and May 2015 for validation, and those after May 2015 for testing. The partitioning of the dataset not only avoids the overlaps between the training, validation, and test datasets, but also helps to evaluate the method’s ability for predicting future unknown viruses based on the previously known viruses. We previously used the data before May 2015 and the partition date January 2014 in \cite{ren2017virfinder}. We updated the dataset to include new viruses after May 2015 and it was natural to use them as test data. 

Since sequences in real metagenomics data are of various lengths ranging from hundreds to thousands of base pairs, we fragmented the genomes into non-overlapping sequences of different sizes, $L=150$, $300$, $500$, $1000$, and $3000$ bp. We built models for sequences of each size, respectively. Table \ref{refdata} shows the numbers of sequences in different sizes for training, validation, and testing data, respectively. The dataset is paired with the same number of prokaryotic sequences, fragmented from RefSeq and partitioned by the exact same dates. The NCBI accession numbers of the viral and prokaryotic RefSeqs can be found in the supplementary materials. 

\small
\begin{table}[]
	\centering
	\caption{The number of viral sequences of various sizes from viral genomes discovered before 2014.1, between 2014.1-2015.5, and after 2015.5. The three parts of the dataset partitioned by dates were used for training, validation, and testing, respectively.
}
	\label{refdata}
	\begin{tabular}{lllll}
		\toprule
		Fragment length & Before 2014.1 & Between 2014.1-2015.5 & After 2015.5 & Total   \\
		 & (training) & (validation) & (test) &    \\
		\midrule
		150 bp          & 505,259                  & 164,918                            & 355,204             & 705,697 \\
		300 bp          & 252,630                  & 82,458                             & 177,416             & 512,504 \\
		500 bp          & 154,640                  & 50,350                             & 106,298             & 311,288 \\
		1000 bp         & 77,014                   & 25,087                             & 52,956              & 155,057 \\
		3000 bp         & 25,263                   & 8,246                              & 17,385              & 50,894 \\
		\bottomrule
	\end{tabular}
\end{table}

\subsection{Predicting viral sequences using convolutional neural networks
}

We used deep learning techniques and developed a powerful framework for predicting viral sequences. Given a query sequence, the framework gives a score between 0 and 1, and the larger score indicates the higher possibility of being a viral sequence. Previously k-mer frequencies were used as features and the logistic Lasso regression model was built \cite{ren2017virfinder}. The success of the method confirmed that virus and their host have different k-mer usage. Those k-mers can be generalized as motifs, which can be described using the position weight matrix (PWM). We expected that using motifs as features could increase the model flexibility and would produce a better prediction model. Thus, we designed a structure of convolutional neural networks (CNN, ConvNet) that extracts motif intensities in sequences and then used them as features for prediction. The advantage of the CNN algorithm is that it simultaneously learns motif patterns and the prediction function from the data, and there is no need to specify motifs beforehand. 

\begin{figure}
	\centering
	\includegraphics[width=1\linewidth]{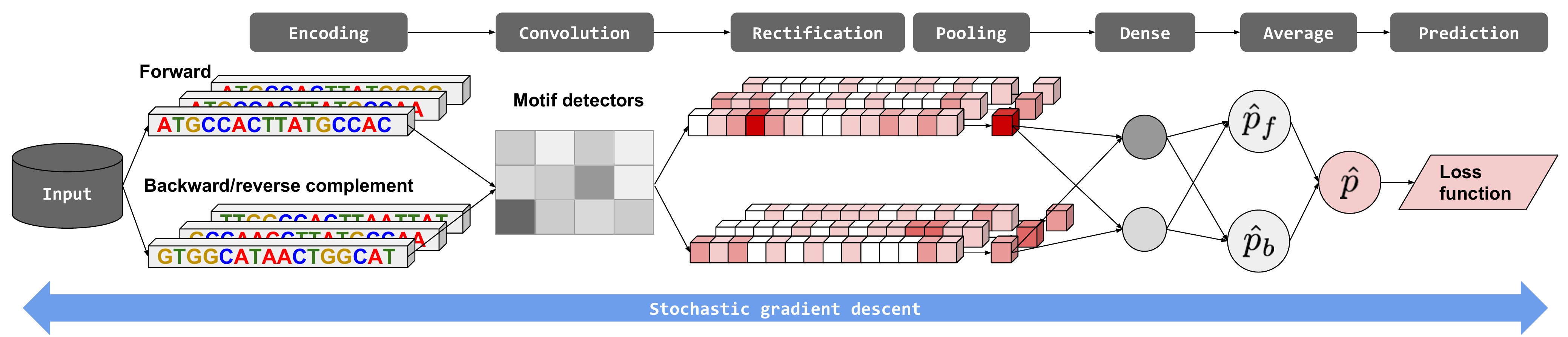}
	\caption{The deep learning framework of DeepVirFinder.}
	\label{fig:framework}
\end{figure}

We call our method, DeepVirFinder. The model consists of a convolutional layer, a pooling layer, a fully connected layer, and several dropout layers (Figure \ref{fig:framework}). We first encoded the DNA sequence of length $L$ using one-hot encoding method with $(A, C, G, T)$ coded as $(1000, 0100, 0010, 0001)$, respectively, resulting in a $4 \times L$ matrix. A rectifier activation convolutional layer contains $M$ motifs of length $f$, and each motif scans the sequence from the beginning to the end obtaining a series of the motif intensities by applying the convolution operation. Thus, the output of the convolutional layer is an $M\times (L-f+1)$ matrix. A max pooling layer reduces the dimension by keeping only the highest intensity for each motif, resulting in an $M\times1$ matrix. A dropout regularization layer of rate 0.1 follows for reducing overfitting in neural networks by randomly dropping a few dimensions. The subsequent layer is a dense layer containing $M$ fully connected neurons with the output matrix of dimension $M\times 1$. With another dropout layer of rate 0.1, the output is summarized using a sigmoid function to generate a prediction score ranging from 0 to 1. Since the DNA sequence is double stranded, and the contigs in real data can come from both strands, the prediction score should be identical no matter the forward strand or the backward strand is provided as the input. Thus, we apply the same network to the reverse complement of the original sequence, and the final prediction score is the average of the predictions from the original and the reverse complement sequences. The same technique was used in \cite{wang2018define, quang2017factornet}. The network is learned by minimizing the binary cross-entropy loss through back-propagation using Adam optimization algorithm for stochastic gradient descent with learning rate 0.001 \cite{kingma2014adam}.

\subsection{Collection of metavirome datasets}

To achieve high prediction accuracy, deep learning algorithm needs a large amount of training data. Though a large number of training sequences were obtained from RefSeq, there is a potential to enlarge the training dataset by including viral sequences from metavirome sequencing data. Metavirome targets at sequencing mainly viruses by removing prokaryotic cells in samples using physical 0.22$\mu m$ filters. Metavirome sequencing does not rely on culturing viruses in the lab so that a wide diversity of viruses can be captured including unknown viruses. A few studies have used this technique to extract viruses and sequenced viral genomes in human gut and ocean \cite{norman2015disease, reyes2015gut, minot2011human, roux2016ecogenomics}. Normal et al. sequenced viruses in the human gut sample from IBD patients using Illumina sequencing technology \cite{norman2015disease}. Reyes et al. studied viruses in fecal samples from Malawian twins with Severe Acute Malnutrition (SAM) using Roche 454 sequencing technology \cite{reyes2015gut}. Minot et al. and Kim et al. investigated virome in health human gut using Roche 454 \cite{minot2011human, kim2016centrifuge}. For marine virome, Tara Ocean Virome project collected the largest number of virome samples from both surface- and deep-ocean viruses over the world \cite{roux2016ecogenomics}. 

We collected the metavirome samples from those studies and aimed to add more viral diversity, especially unknown viruses, to the training data. 
We were careful in quality control of the samples because it is likely that the sample can be contaminated by prokaryotic DNA, since the physical filters may not exclude small sized prokaryotic cells.
The details of preparation of metavirome data and quality control can be found in the supplementary materials and Table \ref{tab:addmeta}. 
Up to 1.3 million of sequences were generated from the metavirome data, and they were the combined with sequences derived from viral RefSeqs before May 2015 for training. The new model was evaluated and compared with the original model trained based on RefSeqs only, using the test sequences from RefSeqs after May 2015.

\subsection{Simulation of metagenomic datasets}

Metagenomic samples were simulated based on species abundance profiles derived from a real human gut metagenomic sample (accession ID SRR061166, Platform: Illumina). 
Given a total budget of base pairs in a sample, the number of base pairs for contigs from each genome was computed proportionally based on the abundance profile. For each reference genome, contigs were sampled randomly and independently from the genome, where the contig length follows the same distribution as that in a real human metagenomics dataset for colorectal carcinoma patients (Figure \ref{fig:simu}a), until the number of base pairs reaches the total budget. The details can be found in the supplementary materials. 

In metagenomic studies, there are two types of sampling strategies. One is referred to as cellular metagnomes that all the genetic materials, including bacteria, archaea, and viruses, are sampled and sequenced. Another type of data is virus fractionated metagenomes (also called as metavirome) where cellular organisms are filtered out before sequencing and mostly viruses are sequenced. 
To mimic the different viral fractions in real metagenomic dat, the abundance profile was rescaled to make samples of three different viral fractions 10\%, 50\%, and 90\%, while keeping the relative abundance within viruses and that within hosts the same. 

We used the model trained by RefSeqs before May 2015 to predict the simulated contigs. For contigs smaller than 300 bp, the model trained by 150 bp was used for prediction; for contigs between 300-500 bp, the 300 bp trained model was used; for contigs between 500-1000 bp, the 500 bp trained model was used, and for contigs >1000 bp, the 1000 bp model was used for prediction. In order to compare scores of contigs across different lengths, we obtained the score distributions for host contigs of different sizes, and converted scores to p-values by comparing the scores with the host score distribution. The area under the ROC curve (AUROC) and the area under the Precision-recall curve (AUPRC) were used to evaluate the prediction performance. 

\subsection{Viral analysis of human gut metagenomics from patients with colorectal cancer}

Human gut metagenomics samples from patients with colorectal carcinoma (CRC) and the control group were downloaded from European Nucleotide Archive (ENA) database (http://www.ebi.ac.uk/ena) with accession number ERP005534. The samples from 53 cancer patients and 61 normal individuals were randomly split into 2/3 for training and 1/3 for testing. The patient ID and the disease status can be found in the supplementary materials. The metagenomics samples from training were combined and cross-assembled. We filtered contigs smaller than 500 bp to guarantee high accuracies in the downstream analysis including viral prediction and contig binning. We used DeepVirFinder to predict viral contigs longer than 500 bp. To control the false discovery rate, the predicted $p$-value for each contig was converted to a $q$-value. The $q$-value is an estimation of the proportion of false prediction if the prediction is made at the level of the corresponding p-value. Contigs were sorted by q-values from the smallest to the largest, and the contigs having q-values <0.01 were predicted as viruses. The viral contigs predicted by DeepVirFinder were then grouped into contig bins, and the abundance of contig bins was derived based on the read mapping results.
To study the association between the viruses and the cancer status, we built a logistic regression classifier with Lasso penalty to predict the CRC status based on the bin abundance on training data, and evaluated the performance on testing data. The details can be found in the supplementary materials.

\section{Results}


\subsection{Determining the optimal model for DeepVirFinder}

In convolutional neuron networks, two critical parameters, the length of motifs (or filters) and the number of motifs, determine the complexity of the model.  To find the best parameter settings, we trained the model with different combinations of the two parameters using the data before January 2014, and evaluated AUROC on the validation dataset. We studied the motif length ranging from 4 to 18 and the number of motifs from 100, 500, 1000, and 1500. We observed that as motif length increased from 4 to 8, the validation AUROC increased rapidly. The highest AUROC achieved when the motif length was around 10, and the value kept in the same level as the motif length future increased (Figure \ref{fig:parameterselection}a, red curves). For example, for the model trained with 500 bp sequences, when fixing the model having 1000 motifs, the validation AUROC increased from 0.7747 to 0.9464 as motif length increased from 4 to 8, and achieved the highest value of 0.9496 when motif length is 10. The trend is similar for all other sequence sizes and numbers of motifs. Thus, we set the motif length as 10 in the final model. Note that the optimal k-mer length is 8 in VirFinder, a similar range as the motif length chosen here. 

We next study the effect of the number of motifs on the model performance by fixing the motif length and increasing the number of motifs from 100 to 1500. The validation AUROC gradually increased with the number of motifs (Figure \ref{fig:parameterselection}b). For example, for the 500 bp model, the validation AUROC was 0.8990, 0.9402, 0.9497, and 0.9500 for models using 100, 500, 1000, and 1500 motifs. The number of neurons in the dense layer had the similar effect on the model performance. Considering the model simplicity and the computational intensity, we chose to use 1000 motifs in the convolutional layer and 1000 neurons in the dense layer in the final model. 

The model is trained based on stochastic gradient descent and back-propagation. The weights in the model are updated by iteratively passing the entire training data forward and backward. One iteration is called one epoch. Using more epochs produces higher training accuracy but it can also cause overfitting. We observed the validation AUROC increased quickly in the first 20 epochs and was stabilized after 30 epochs (Figure \ref{fig:parameterselection}c). Thus we use 30 epochs in the final model. 

Comparing models across different sequence lengths, we observed that longer sequences have higher prediction accuracy. For example, the models with 1000 motifs of 10 bp and trained using 30 epochs have validation AUROCs of 0.8635, 0.9210, 0.9496, 0.9668, 0.9784 for 150, 300, 500, 1000, and 3000 bp sequences. Longer sequences contain more information and thus are easier to make predictions. For sequences at least 300 bp, the corresponding AUROC on training dataset were all above 0.98, indicating the models were fully trained. The low AUROC for the model of 150 bp sequences was due to the generic/intrinsic difficulty of predicting the very short sequences. 

Once the parameters were determined, we trained the model using all sequences before May 2015 (training plus validation datasets). We evaluated model performance on sequences after May 2015, independent from all the training and parameter tuning process, to obtain the unbiased evaluation of the model performance.

\subsection{DeepVirFinder outperforms VirFinder}

\begin{figure}
	\centering
	\includegraphics[width=1\linewidth]{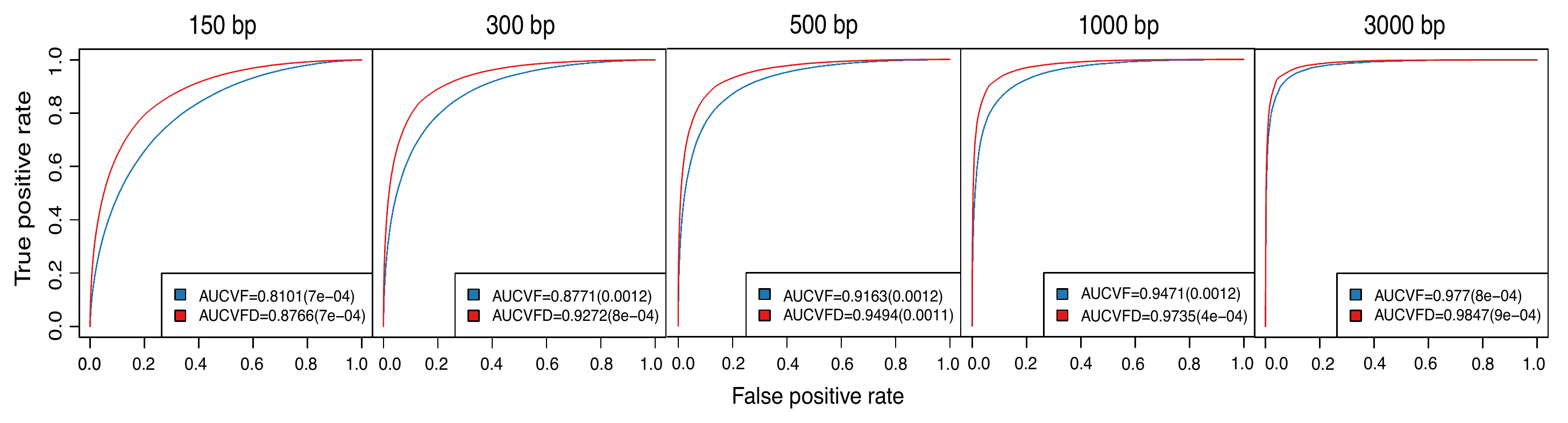}
	\caption{ROC and AUROC for VirFinder and DeepVirFinder when trained on sequences before May 2015, and tested on sequences after May 2015. The numbers in brackets are the standard errors estimated using 30 bootstrap samples.}
	\label{fig:vf_vfd}
\end{figure}

We compared the performance of the new model DeepVirFinder with that of VirFinder \cite{ren2017virfinder}. VirFinder used k-mer frequencies as features for sequences and trained a regularized logistic regression model to predict viral sequences. To make a fair comparison, both methods were trained using the sequences before May 2015 and assessed on data after May 2015. DeepVirFinder outperformed VirFinder at all sequence lengths, where the ROC curve for DeepVirFinder is always higher than that for VirFinder (Figure \ref{fig:vf_vfd}). The improvement in AUROC is more remarkable for short sequences of length <1000 bp. For example, DeepVirFinder had AUROC of 0.8766, 0.9272, and 0.9494 for 150, 300, and 500 bp sequences, while the corresponding AUROC scores for VirFinder were 0.8101, 0.8771, and 0.9163, reflecting 8.2\%, 5.7\%, and 3.6\% increase, respectively. For 1000 bp sequences, DeepVirFinder improved the AUROC from 0.9471 to 0.9735 (2.8\% increase), and for sequence of size 3000 bp, the increase from 0.977 to 0.9847 was minimal but still significant ($p$-value for one-sided t-test, 5.896e-16). 
With the improvement, the new model can predict sequences as short as 300 bp with a decent accuracy (AUROC 0.9272), which is at the similar level of AUROC for 500 bp in the previous model (AUROC 0.9163). 

\subsection{Predicting sequences of various lengths using different models and the effect of mutations }

\begin{figure}
	\centering
	\includegraphics[width=1\linewidth]{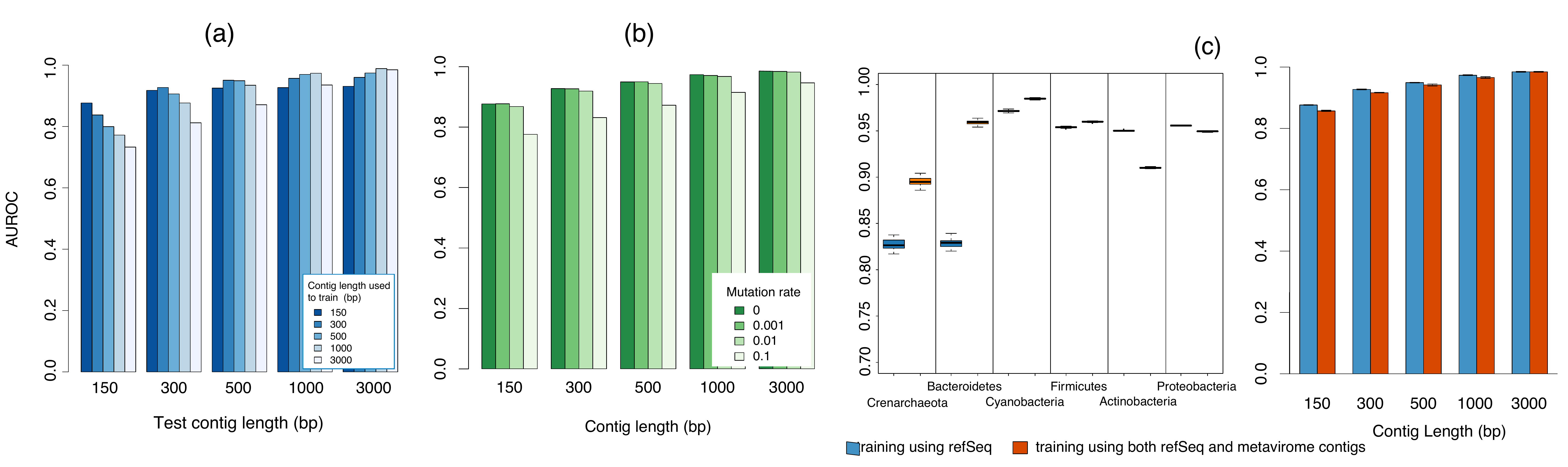}
	\caption{(a) AUROC for different combinaions of contig lengths used for training and testing; (b) AUROC for prediction when adding mutations at different rates; (c) Comparison of AUROC between the model trained using only viral refSeqs, and the model trained using the enlarged dataset including metavirome contigs. Left: the overall AUROCs between the two models at different contig sizes. Right: the AUROCs for viruses from different host phyla. The $*$ is for statistical signidicance with the $p$-value for one-sided t-test smaller than <2.2e-16.}
	\label{fig:traintestdifferrorrate}
\end{figure}

We trained models using fixed sized sequences, while real metagenomics contigs are of different lengths. Given a query sequence for prediction, a natural question is which model we should use for prediction. We evaluated the performance of different models for predicting sequences of different lengths. For example, we predicted 150 bp sequences after May 2015 using models trained by 150, 300, 500, 1000, 3000 bp sequences before May 2015, respectively. The highest AUROC achieves when using the model trained by 150 bp sequences (Figure \ref{fig:traintestdifferrorrate}a). Similarly, using the model trained by 300 bp sequences had the best performance for predicting 300 bp sequences, and the same conclusion holds for 500 and 1000 bp sequences. For sequences of length > 1000, there was an invisible difference between models. Therefore, we decided to use the 150 bp model for predicting sequences <300 bp, the 300 bp model for sequences of the size between 300 to 500 bp, 500 bp model for sequences between 500 and 1000 bp, and use the 1000 bp model to predict sequences greater than 1000 bp.

Since viruses evolve at a higher mutation rate than bacteria, we test how the models tolerate mutations. This also served as a test of the sensitivity of the models to sequencing errors. For the sequences in the testing dataset, we randomly introduced mutations by replacing the original letter at each position by another letter with equal probability at the rate of (0.001, 0.01, 0.1). We compare the AUROC for the mutated sequences with that of no mutations. Within each contig size, the AUROC score drop less than 0.06\% at 0.001 mutation rate, 0.66\% at 0.01 mutation rate, and 7.96\% at 0.1 mutation rate (Figure \ref{fig:traintestdifferrorrate}b). Thus, the deep learning models are not sensitive to the viral mutation rate of $\leq$0.001 as suggested in \cite{lauring2013role} as well as the sequencing error rate of 0.001 \cite{glenn2011field}.

\subsection{Enlarging the training dataset by adding millions of metavirome contigs}

Though a large number of training sequences were obtained from viral RefSeqs, viral sequences from metavirome datasets could be added to the training to boost the prediction accuracy of deep learning algorithm. We collected a large set of metavirome samples from several large scaled studies, and we carefully selected the samples that had high quality to reduce the possibility of contamination of prokaryotic DNA. Reads from those samples were assembled and the resulting millions of viral contigs were used to generate more viral sequences for training. 
The new model was evaluated using the test sequences from RefSeqs after May 2015, and compared with the original model trained based only on RefSeqs.

We investigated the AUROC for predicting viruses from different host phyla. 
The new model trained using the enlarged dataset had significantly higher AUROC scores for the viral groups that are under-represented in NCBI RefSeq database, compared with the original model with all $p$-values for one-sided t-test $<$2.2e-16 (Figure \ref{fig:traintestdifferrorrate}c,left).
For example, only 2.08\% viruses in RefSeqs infect Bacteroidetes. Because of the low representation, the original model trained using only RefSeqs had a relatively low AUROC of 0.8287 for predicting viruses infecting Bacteroidetes. After adding the metavirome contigs, the AUROC was improved to 0.9591, with a large 0.1304 increase. 
Similarly, the new model improved the AUROC from 0.8272 to 0.8952 for viruses infecting Crenarchaeota, and from 0.9714 to 0.9847 for viruses infecting Cyanobacteria, which only represent 2.23\% and 4.39\% of the viruses in RefSeq database, respectively. 
The results suggest the distribution of different viruses in RefSeq database is different from the viral community in the environment. This can be due to the fact that the viruses in RefSeq mostly were obtained by cultivation in lab so that only viruses whose hosts are cultivable can be isolated. 
Therefore, adding the contigs from the metavirome dataset reduced the bias and improved the prediction accuracy for the viruses under-represented in the RefSeqs. 

We also noticed that the viruses infecting Proteobacteria and Actinobacteria, the two most abundant virus types taking up to 63\% in RefSeqs, had an decreased AUROC when used the enlarged dataset for training. 
This may also due to the difference of the virus type distributions between the original RefSeq training data and the enlarged training data.
The above results were for 500 bp contigs, and the conclusion holds for other sequence lengths (data not shown). 
Due to the decrease of AUROC decresing for the two major viral groups, the overall AUROC was also slightly decreased for the new model (Figure \ref{fig:traintestdifferrorrate}c,right).
Thus, we suggest to use the new model trained with the enlarged data to predict viruses from under-represented groups, and use the original model if the viruses are mainly from the common groups.

Though we have used strict quality control when choosing the datasets, it is still possible that the sequences generated from metavirome datasets contain a small portion of host genomes. Besides, the assembly may cause chimeric contigs, either composed by a mixture of viral and host reads, or a mixture of different viruses. 
Those factors may introduce more or less contaminations to the training data that is hardly avoidable. 
On the other hand, we assess the model based on sequences from RefSeqs after May 2015, which has a similar distribution with that for Refseqs before May 2015. 
We expect the model trained with the enlarged dataset will have better performance if the testing data was from the true viral distribution. 

%


\subsection{Evaluation of DeepVirFinder on simulated metagenomic contigs of variable lengths}

\begin{figure}
	\centering
	\includegraphics[width=0.95\linewidth]{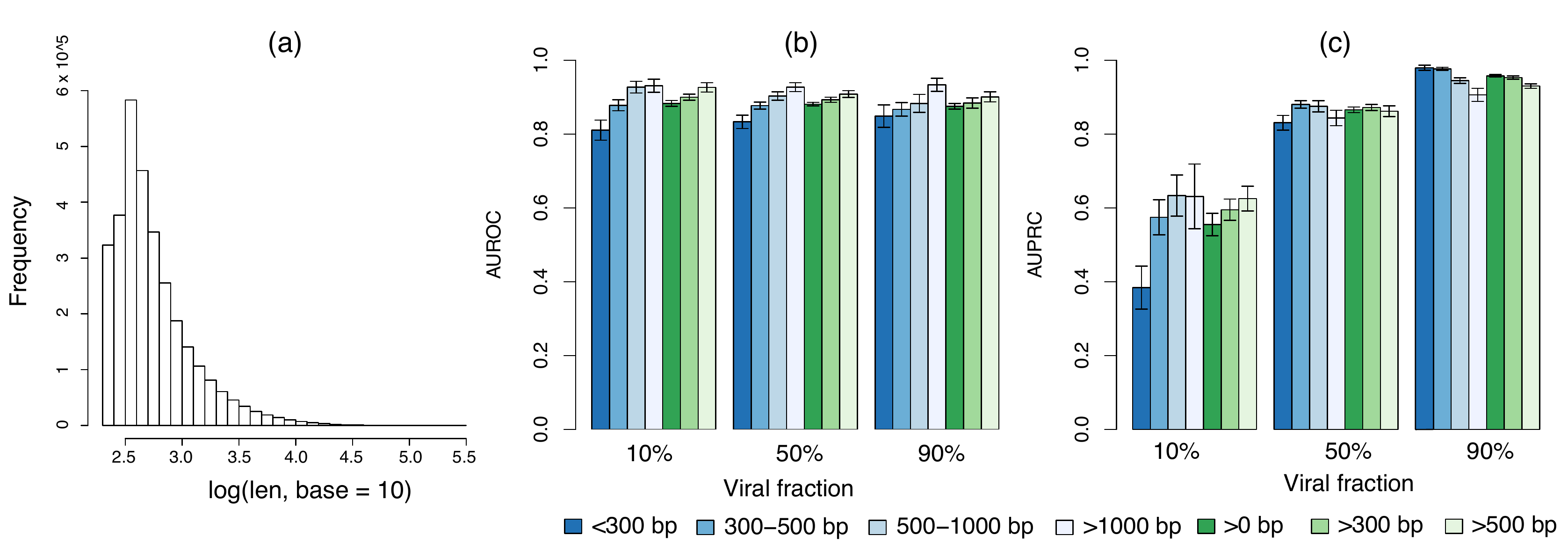}
	\caption{(a) The distribution of contig length used for simulating metagenomic samples, and the AUROC (b) and AUPRC (c) for predicting viral sequences with various viral fractions for different sizes of contigs.}
	\label{fig:simu}
\end{figure}

To test the performance of DeepVirFinder on predicting viral contigs in metagenomics data, we simulated several metagenomics samples based on the abundance profile of a real human gut metagenomic sample,  and evaluated the model accuracy for identifying viral contigs in the simulated samples. The simulated contig was of the size between hundreds of base pairs to thousands of base pairs, with the majority ranging between 300-1000 bp. We used different models (trained with RefSeqs) to predict contigs of different sizes. Both AUROC and AUPRC were used to measure the prediction performance.  In general, AUROC increases as the contig length increases, having the same trend as in Figure \ref{fig:vf_vfd}. For example, the AUROC scores for contigs of length <300 bp, 300-500 bp, 500-1000 bp, and >1000 bp were about 0.8317, 0.8767, 0.8966, and 0.9451 on average. When predicting contigs of length across multiple intervals, AUROC is 0.8829 for all contigs, 0.8952 for contigs >300 bp, and 0.9129 for contigs >500 bp (Figure \ref{fig:simu}a). Thus, in the real data application, we predict contigs >300 bp in order to control the overall AUROC ~0.90.

Different viral fraction does not markedly affect AUROC, since the true positive rate and the false positive rate are defined based only on the relative rate within the viral group and the host group independently. On the other hand, AUPRC is strongly affected by the viral fraction, since the precision depends on the ratio between viral and host fractions. For example, the AUPRC for contigs of length >500 bp is 0.9296 for the sample with 90\% viral fraction, while that is 0.8638 and 0.6437 for samples with 50\% and 10\% viral fractions, respectively (Figure \ref{fig:simu}b). AURPC had large variations for the sample with 10\% viral fraction, compared to that for the sample with 50\% and 90\% viral fractions. This is due to the small number of viruses in the 10\% viral fraction sample.

\subsection{Identifying viruses in the gut microbiome associated with colorectal cancer}

Though the effect of bacteria on the CRC has been studied \cite{feng2015gut,vogtmann2016colorectal, nakatsu2015gut,nakatsu2015gut}, the association between viruses and the disease have not been investigated yet. 
Since viruses play important roles in controlling host population and altering host metabolism, it is of significance to investigate the gut viruses in patients with CRC.   
We used DeepVirFinder to identify viruses in the gut microbiome, and ask if any virus is associated with the disease status.  

The metagenomic samples are cross assembled, resulting in 1,335,046 contigs of length greater than 500 bp. DeepVirFinder identified 51,138 viral contigs and the false discovery rate was controlled at the rate 0.01. Those contigs were then grouped into 175 contig bins. Using the average RPKM as the feature for each bin and the disease status as the response variable, we built a logistic Lasso regression classifier to predict cancer status based on the viral bins. The AUROC score was 0.7557, and 10 viral bins selected by the Lasso classifier were associated with the CRC status (Table \ref{tab:bin}). Six of the bins had positive coeficients (high abundance in CRC) and four bins had negative coefficients (high abundance in controls). This implies the great potential of using gut viruses as an non-invasive diagnosis method for CRC. 

Comparing the sequences in the bins to NCBI database, six bins had similar sequences to known viruses. Most of the viruses infect bacteria (called phages), while we also noticed one endogenous virus infecting human cells. 
In addition, Bin7, with a negative coeficient -0.0193 in the regression model, contains sequences similar to crAssphage. This is consistent with the fact that crAssphage is a highly abundant virus in health gut \cite{dutilh2014highly}. 
The rest four bins with no similarity to known viruses are newly discovered viruses. 
We also searched the proteins in contigs against Pfam database. 
On average 65.68\% of contigs contained proteins, showing DeepVirFinder's ability of identifying virus sequences from non-coding regions over the other gene homology-based method.  
Seven of the 10 bins contained phage associated proteins, such as tail, capsid, integrase, connector, holin, and portal, indicating those bins were trully from viruses. 
In addition, all bins except Bin188 had proteins with domains of unknown function (DUF). 
This also indicates those bins were most likely to be viruses because viral proteins are less characterized than bacteria in Pfam database \cite{roux2015virsorter}, and in fact this was a criterion used for viral prediction in Roux et al. \cite{roux2015virsorter}


\section{Conclusions}

Viruses play crucial roles in regulating microbial communities, but our knowledge of viruses has been delayed by the tradition experiment techniques and the computational methods for a long time.  
We developed a powerful method, DeepVirFinder, to identify viral sequences in metagenomic data. 
To the best of our knowledge, it is the first time that deep learning techniques are used for problems in metagenomics. 
Built based on the deep learning techniques, and trained with a large number of viral sequences, DeepVirFinder outperformed the the state-of-the-art method, VirFinder at all contig lengths. 
Enlarging the training data with additional million of viral sequences from the environmental samples futher improved the prediction accuracy for the under-represented viral groups. 
We applied DeepVirFinder to human gut metagenomic samples and identified a group of key viruses associated with CRC. 
As the numbers of RefSeqs and environmental metavirome samples keep increasing, we expect DeepVirFinder will keep increasing its power in identifying viruses in metagenomic data. 
In future studies, it will be interesting to compare the motifs learned from the deep learning model with the real biological motifs, though it is not trivial to collect the dataset of viral motifs. 
Overall, DeepVirFinder significantly improves the precision and recall rates of identifying viruses in metagenomic data, and it will greatly accelerate the discovery rate of viruses.

\subsubsection*{Acknowledgments}
We thank Drs. Michael S. Waterman, Gesine Reinert, Ying Wang, Rui Jiang, Yang Lu, Mr. Weili Wang, and Mr. Luigi Manna for helpful discussions and suggestions.
The research was supported by the U.S. National Institutes of Health R01GM120624, National Science Foundation DMS-1518001, and National Natural Science Foundation of China (11701546).
This research utilized resources of HPC (High-Performance Computing), which is supported by the University of Southern California.



\bibliographystyle{plainnat}
\small
\bibliography{reference}

\newpage
\pagebreak
\begin{center}
	\textbf{\large Supplemental Materials: Identifying viruses from metagenomic data by deep learning}
\end{center}
\setcounter{equation}{0}
\setcounter{figure}{0}
\setcounter{table}{0}
\setcounter{page}{1}
\makeatletter
\renewcommand{\theequation}{S\arabic{equation}}
\renewcommand{\thefigure}{S\arabic{figure}}
\renewcommand{\thetable}{S\arabic{table}}
\renewcommand{\bibnumfmt}[1]{[S#1]}
\renewcommand{\citenumfont}[1]{S#1}

\paragraph{Enlarging training data with metavirome contigs}

To measure the contamination rate of each sample, we mapped reads using bowtie2 (2.3.2) \cite{langmead2012fast} in the sample to a reference database containing 7726 prokaryotic RefSeqs used in VirSorter \cite{roux2015virsorter} and human RefSeq (GRCh38.p7), and used the mapping rate as the contamination rate. In order to reduce the effect of prophages inserted in prokaryotic genomes on the measurement, the prophages were detected using VirSorter and removed from the genomes. 
Samples with contamination rate lower than 5\% were considered in the next step. The samples from the same sequencing platform were combined and cross-assembled into contigs. Megahit \cite{li2015megahit} was used to assemble data from Illumina sequencing, and Ray \cite{boisvert2010ray, boisvert2012ray} was used to assemble reads from Roche 454. To further reduce the non-viral contamination \textit{in silico}, the resulting contigs were filtered through VirSorter and only the viral contigs with the highest confidence (Category I and II) were used for training. Samples from TOV were carefully cross-assembled and filtered in Roux et al. \cite{roux2016ecogenomics}, so the resulting viral contigs were directly used for training in our study. The metavirome contigs were then fragmented into millions of fixed-length sequences. Table \ref{tab:addmeta} lists the details of the metavirome datasets. The metavirome sequences were combined with sequences derived from viral RefSeqs before May 2015 for training. The new model was evaluated and compared with the original model trained based on only RefSeqs, using the test sequences from RefSeqs after May 2015.

\paragraph{Simulation of metagenomic datasets}

Metagenomic samples were simulated based on species abundance profiles derived from a real human gut metagenomic sample (accession ID SRR061166, Platform: Illumina) from the Human Microbiome Project (HMP) \cite{peterson2009nih}, commonly used for metagenomic data analysis \cite{boisvert2012ray, luo2014mytaxa, rampelli2016viromescan, brittnacher2016gutss}. 
We first mapped reads from sample SRR061166 using bwa-0.7.15 \cite{li2009fast} to virus and host genomes sequenced after May 2015 to generate the abundance profile. Here we only used RefSeqs after May 2015 for evaluation to avoid any overlap with the training dataset, i.e. RefSeqs before May 2015. Following a similar procedure as in VirFinder, reads from each sample were first mapped to viral RefSeqs and the remaining unmapped reads were then mapped to host RefSeqs using the command of “bwa mem”. About 2\% of reads can be mapped to viral genomes, lower than the range of previously estimated viral fraction 4-17\% for human gut metagenomics. This is largely due to the incomplete viral RefSeqs genomes used here. The abundance profiles can be found in the supplementary materials. 

We simulated metagenomic contigs based on the abundance profile. Given a total budget of 10 million base pairs for contigs, the number of base pairs for contigs from each genome was computed proportionally. For each reference genome, contigs were sampled randomly and independently from the genome, where the contig length follows the same distribution as that in a real human metagenomics dataset for colorectal carcinoma patients, until the number of base pairs reaches 10 Mbp. Note that here we generated contigs directly from genomes, instead of directly simulating reads and then assembling contigs. We avoided chimeric contigs that were artificially assembled using reads from different genomes. The R packages “ROCR” and “caTools” were used to compute AUROC and AUPRC, and the variation were evaluated using 30 bootstrap samples.

\paragraph{Viral analysis of human gut metagenomics from patients with colorectal cancer}

Human gut metagenomics samples from patients with colorectal cancer and the control group were downloaded from European Nucleotide Archive (ENA) database (http://www.ebi.ac.uk/ena) with accession number ERP005534. The samples from 53 cancer patients and 61 normal patients were randomly split into 2/3 for training and 1/3 for testing. The patient ID and the disease status can be found in the supplementary materials. The metagenomics samples from training were combined and cross-assembled using Megahit \cite{li2015megahit} in default settings. The majority 64\% of the assembled contigs have the length ranging from 300-1000 bp (Figure \ref{fig:simu}a). We filtered contigs smaller than 500 bp to guarantee high accuracies in the downstream analysis including viral prediction and contig binning. We used DeepVirFinder to predict viral contigs longer than 500 bp. To control the false discovery rate, the predicted p-value for each contig was converted to a q-value using the R package “qvalue” \cite{storey2003positive}. The q-value is an estimation of the proportion of false prediction if the prediction is made at the level of the corresponding p-value. Contigs were sorted by q-values from the smallest to the largest, and the contigs having q-values <0.01 were predicted as viruses. The viral contigs predicted by DeepVirFinder were then grouped into contig bins using the software COCACOLA \cite{lu2017cocacola} in the default mode.

To study the association between the viruses and the cancer status, we mapped reads in each sample against the viral contigs using “bowtie2-2.3.2”. The number of reads per kilobase of the contig per million mapped reads (RPKM) was used as a measure of contig abundance, and the average of the RPKM of contigs in each bin was defined as the contig bin abundance. Based on the abundance of viral bins in training data, a logistic regression classifier with L1 penalty was built to predict the cancer status using the R package “glmnet”. The parameter lambda was determined using 5 fold cross-validation. PfamScan and Blastn-2.6.0 (Evalue <1e-5) were used respectively to search proteins and DNA sequence against Pfam and NCBI databases.

\paragraph{Other supplementary tables}
The NCBI accession numbers of the viral and prokaryotic RefSeqs can be found at \href{https://drive.google.com/open?id=1FKXTVCEN2d0-xHnpzUSea7lOjOs9YDDW}{https://drive.google.com/open?id=1FKXTVCEN2d0-xHnpzUSea7lOjOs9YDDW}.

The species abundance profile for simulated metagenomic samples can be found at \href{https://drive.google.com/open?id=1lM4be7vsUQ-MOnUFl2loG-GgXKqC0GS9}{https://drive.google.com/open?id=1lM4be7vsUQ-MOnUFl2loG-GgXKqC0GS9}.

The sample ID used for identifying viruses in CRC patients can be found at \href{https://drive.google.com/open?id=1Tet1LZfLAS0-ujqvclYztL1Y3rHfklPA}{https://drive.google.com/open?id=1Tet1LZfLAS0-ujqvclYztL1Y3rHfklPA}.

\clearpage

\begin{figure}
	\centering
	\includegraphics[width=1\linewidth]{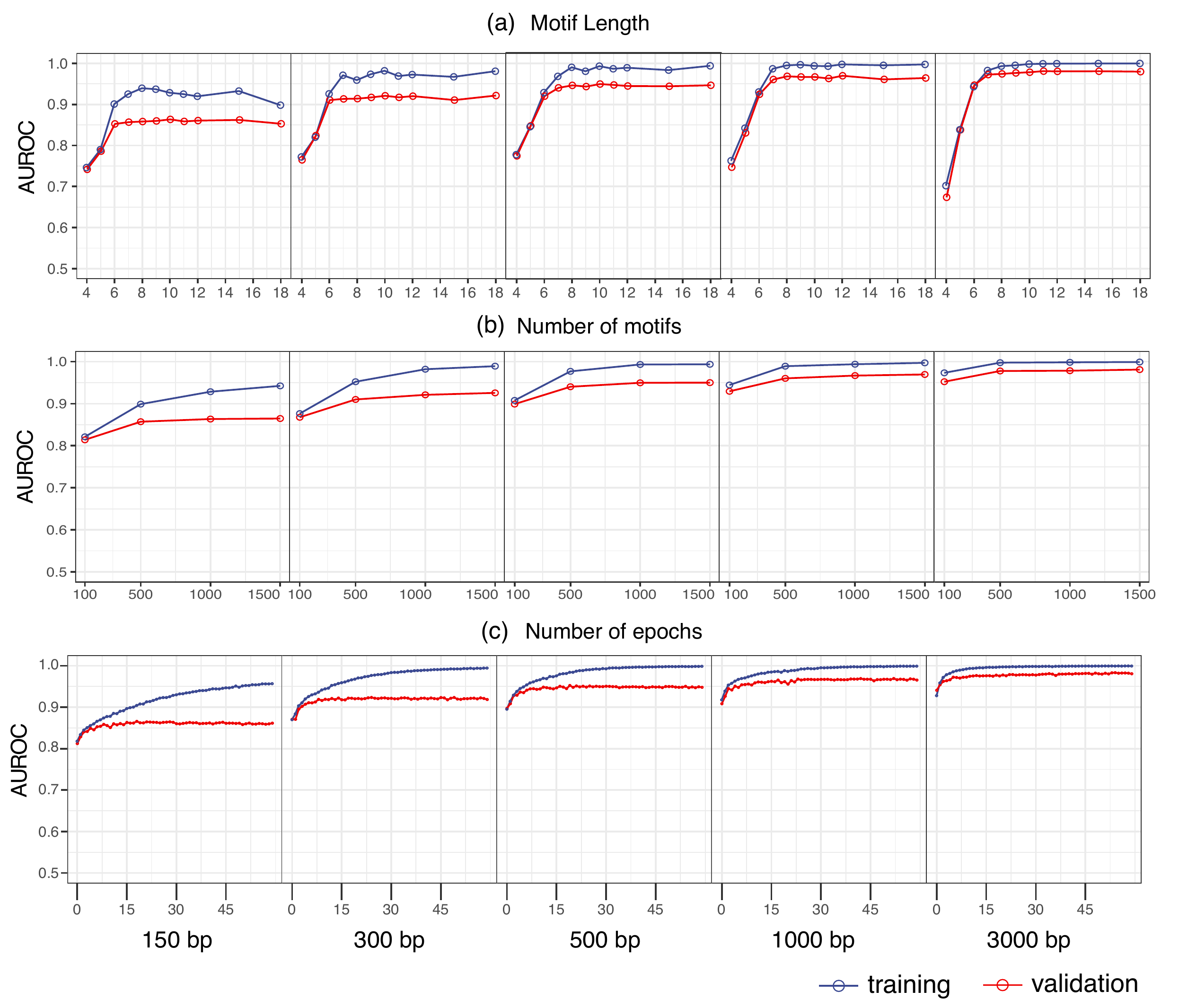}
	\caption{The effect of motif length, number of motifs, and the number of epochs on validation AUROC.}
	\label{fig:parameterselection}
\end{figure}


\begin{table}[]
	\centering
	\caption{(a) Metavirome datasets used for generating more viral contigs for training. (b) Number of sequences at different lengths for each type of metavirome dataset. PE: paired-end reads; \# pure sample: number of samples have contamination rate $<$5\%; Data size: the total size of the data. \\
		\footnotesize{$^*$ Roux et al. \cite{roux2016ecogenomics} used a customized cross-assemble pipeline with MOCAT \cite{kultima2012mocat} and Idba\_ud \cite{peng2010idba, peng2012idba} to assemble the 104 samples from TOV project, and resulting contigs were further filtered by VirSorter to identify 298,383 viral contigs.}}
	\label{tab:addmeta}
	\begin{subtable}[t]{1\linewidth}
		\subcaption{Metavirome datasets}
		\begin{tabular}{p{2cm}p{1.5cm}p{0.2cm}p{1.5cm}p{2cm}p{1cm}p{0.8cm}p{1.5cm}}
			\toprule
			\multicolumn{8}{c}{Human Gut Metavirome} \\
			\hline
			Dataset               & Platform  & PE & \# Sample & \# Pure Sample & Purity Rate & Data Size & Assmebler         \\
			\midrule
			IBD \cite{norman2015disease}    & Illumina  & Y           & 171       & 131                                   & 76.61\%     & 174G    & Megahit     \\
			SAM\cite{reyes2015gut}     & Roche 454 & N           & 320       & 281                                   & 87.81\%     &   &        \\
			Healthy \cite{minot2011human}  & Roche 454 & N           & 18        & 1                                     & 5.56\%      &     8.82G       &       Ray            \\
			Healthy \cite{kim2016centrifuge}    & Roche 454 & N           & 6         & 5                                     & 83.33\%     &           &                   \\ \hline
			\multicolumn{8}{c}{Tara Ocean Metavirome} \\ \hline
			Roux et al (TOV) \cite{roux2016ecogenomics}            & Illumina  & Y           & 104       & NA$^*$                                   & NA$^*$         & 925G    & Customized \\
			\bottomrule
		\end{tabular}
	\end{subtable}
	\newline
	
	\begin{subtable}[t]{1\linewidth}
		\subcaption{Number of sequences in metavirome datasets}
		\begin{tabular}{lllll}
			\toprule
			Fragment length & Human Gut (Roche 454) & Human Gut (Illumina) & TOV       & Total     \\
			& (Roche 454) & (Illumina) &        &      \\
			\midrule
			150 bp          & 9,467                 & 3,389                & 1,354,007 & 1,366,863 \\
			300 bp          & 4,688                 & 1,658                & 671,331   & 677,677   \\
			500 bp          & 2,772                 & 965                  & 398,342   & 402,079   \\
			1000 bp         & 1,337                 & 442                  & 193,373   & 195,152   \\
			3000 bp         & 388                   & 91                   & 56,698    & 57,177   \\
			\bottomrule
		\end{tabular}
	\end{subtable}
\end{table}


\begin{table}[]
	\centering
	\caption{The viral contig bins associated with the CRC. For each bin, the following information is shown: the coeficient of the bin, the number of contigs in the bin, the percentage of contigs containing proteins, the top Pfam hits of the proteins in the bin, and the top BLAST hits to the NCBI database. }
	\label{tab:bin}
	\begin{tabular}{p{1cm}p{1.3cm}p{1.3cm}p{1.2cm}p{4cm}p{3.5cm}}
		\toprule
		Bin ID & Coeficient & \# contigs & protein\% & Pfam hit & Blastn hit \\
		\midrule
		B7  &  -0.0193 & 957 & 61\% & Phage related (tail, capsid, integrase, connector, holin, portal), Tail\_P2\_I, PhageMin\_Tail, Podovirus\_Gp16, 
		RNA\_lig\_T4\, Terminase, DUF  & crAssphage, Unidentified phage \\
		B19 &  0.1029 & 963 & 63.66\% & Phage related (tail, capsid, integrase, connector, holin, portal), 
		TMP-TENI, T4SS, Terminase, DUF  & Escherichia virus, Enterobacteria phage, Salmonella phage, Stx1/Stx2 converting phage, Bacteriophage, Lambda genome, Unidentified phage\\
		B20 &  -0.0743 &  254  & 31.10\% & Phage related (tail, capsid, integrase, connector, holin, portal), 
		TMP (Prophage tail length tape measure protein ), DUF & Homo sapiens isolate endogenous virus, Bacteriophage, Enterobacteria phage, Stx1/Stx2 converting phage, Escherichia phage, Salmonella phage \\
		B60 & 0.1666 &  384 & 53.65\% & Phage related (tail, capsid, integrase, connector, holin, portal), Podovirus, Terminase, DUF & Staphylococcus phage, Unidentified phage, Aureococcus anophagefferen, \\
		B61 & 0.0924 & 702 &  51.85\% & Phage related (tail, capsid, integrase, connector, holin, portal), ADH\_zinc\_N, Terminase, DUF & Streptococcus phage, Unidentified phage clone, Faecalibacterium phage
		\\
		B87 & -0.0223 & 26 &  96.15\% & DUF & NA \\
		B110 & -0.3475 & 72 & 94.44\% & DUF & NA \\
		B188 &  0.0174 & 56 & 35.71\% & Phage\_T4\_gp19 & NA \\
		B218  &  0.0455 & 107 & 85.05\% & Phage\_sheath, 
		DUF  & NA \\
		B227 & 0.1764 & 159 & 84.28\% & DUF & Moraxella phage\\
		\bottomrule
	\end{tabular}
\end{table}

\clearpage
\small
\bibliographystyle{plain}
\bibliography{reference}

\begin{thebibliography}{49}
\providecommand{\natexlab}[1]{#1}
\providecommand{\url}[1]{\texttt{#1}}
\expandafter\ifx\csname urlstyle\endcsname\relax
  \providecommand{\doi}[1]{doi: #1}\else
  \providecommand{\doi}{doi: \begingroup \urlstyle{rm}\Url}\fi

\bibitem[Alipanahi et~al.(2015)Alipanahi, Delong, Weirauch, and
  Frey]{alipanahi2015predicting}
Babak Alipanahi, Andrew Delong, Matthew~T Weirauch, and Brendan~J Frey.
\newblock Predicting the sequence specificities of dna-and rna-binding proteins
  by deep learning.
\newblock \emph{Nature biotechnology}, 33\penalty0 (8):\penalty0 831, 2015.

\bibitem[Boisvert et~al.(2010)Boisvert, Laviolette, and
  Corbeil]{boisvert2010ray}
S{\'e}bastien Boisvert, Fran{\c{c}}ois Laviolette, and Jacques Corbeil.
\newblock Ray: simultaneous assembly of reads from a mix of high-throughput
  sequencing technologies.
\newblock \emph{Journal of computational biology}, 17\penalty0 (11):\penalty0
  1519--1533, 2010.

\bibitem[Boisvert et~al.(2012)Boisvert, Raymond, Godzaridis, Laviolette, and
  Corbeil]{boisvert2012ray}
S{\'e}bastien Boisvert, Fr{\'e}d{\'e}ric Raymond, {\'E}l{\'e}nie Godzaridis,
  Fran{\c{c}}ois Laviolette, and Jacques Corbeil.
\newblock Ray meta: scalable de novo metagenome assembly and profiling.
\newblock \emph{Genome biology}, 13\penalty0 (12):\penalty0 R122, 2012.

\bibitem[Brittnacher et~al.(2016)Brittnacher, Heltshe, Hayden, Radey, Weiss,
  Damman, Zisman, Suskind, and Miller]{brittnacher2016gutss}
Mitchell~J Brittnacher, Sonya~L Heltshe, Hillary~S Hayden, Matthew~C Radey,
  Eli~J Weiss, Christopher~J Damman, Timothy~L Zisman, David~L Suskind, and
  Samuel~I Miller.
\newblock Gutss: An alignment-free sequence comparison method for use in human
  intestinal microbiome and fecal microbiota transplantation analysis.
\newblock \emph{PloS one}, 11\penalty0 (7):\penalty0 e0158897, 2016.

\bibitem[Buchfink et~al.(2015)Buchfink, Xie, and Huson]{buchfink2015fast}
Benjamin Buchfink, Chao Xie, and Daniel~H Huson.
\newblock Fast and sensitive protein alignment using diamond.
\newblock \emph{Nature methods}, 12\penalty0 (1):\penalty0 59--60, 2015.

\bibitem[Dutilh et~al.(2014)Dutilh, Cassman, McNair, Sanchez, Silva, Boling,
  Barr, Speth, Seguritan, Aziz, et~al.]{dutilh2014highly}
Bas~E Dutilh, Noriko Cassman, Katelyn McNair, Savannah~E Sanchez, Genivaldo~GZ
  Silva, Lance Boling, Jeremy~J Barr, Daan~R Speth, Victor Seguritan, Ramy~K
  Aziz, et~al.
\newblock A highly abundant bacteriophage discovered in the unknown sequences
  of human faecal metagenomes.
\newblock \emph{Nature communications}, 5, 2014.

\bibitem[Feng et~al.(2015)Feng, Liang, Jia, Stadlmayr, Tang, Lan, Zhang, Xia,
  Xu, Jie, et~al.]{feng2015gut}
Qiang Feng, Suisha Liang, Huijue Jia, Andreas Stadlmayr, Longqing Tang, Zhou
  Lan, Dongya Zhang, Huihua Xia, Xiaoying Xu, Zhuye Jie, et~al.
\newblock Gut microbiome development along the colorectal adenoma--carcinoma
  sequence.
\newblock \emph{Nature communications}, 6:\penalty0 6528, 2015.

\bibitem[Glenn(2011)]{glenn2011field}
Travis~C Glenn.
\newblock Field guide to next-generation dna sequencers.
\newblock \emph{Molecular ecology resources}, 11\penalty0 (5):\penalty0
  759--769, 2011.

\bibitem[Kelley et~al.(2016)Kelley, Snoek, and Rinn]{kelley2016basset}
David~R Kelley, Jasper Snoek, and John~L Rinn.
\newblock Basset: learning the regulatory code of the accessible genome with
  deep convolutional neural networks.
\newblock \emph{Genome research}, 26\penalty0 (7):\penalty0 990--999, 2016.

\bibitem[Kim et~al.(2016)Kim, Song, Breitwieser, and
  Salzberg]{kim2016centrifuge}
Daehwan Kim, Li~Song, Florian~P Breitwieser, and Steven~L Salzberg.
\newblock Centrifuge: rapid and sensitive classification of metagenomic
  sequences.
\newblock \emph{Genome research}, 26\penalty0 (12):\penalty0 1721--1729, 2016.

\bibitem[Kingma and Ba(2014)]{kingma2014adam}
Diederik~P Kingma and Jimmy Ba.
\newblock Adam: A method for stochastic optimization.
\newblock \emph{arXiv preprint arXiv:1412.6980}, 2014.

\bibitem[Kultima et~al.(2012)Kultima, Sunagawa, Li, Chen, Chen, Mende,
  Arumugam, Pan, Liu, Qin, et~al.]{kultima2012mocat}
Jens~Roat Kultima, Shinichi Sunagawa, Junhua Li, Weineng Chen, Hua Chen,
  Daniel~R Mende, Manimozhiyan Arumugam, Qi~Pan, Binghang Liu, Junjie Qin,
  et~al.
\newblock Mocat: a metagenomics assembly and gene prediction toolkit.
\newblock \emph{PloS one}, 7\penalty0 (10):\penalty0 e47656, 2012.

\bibitem[Langmead and Salzberg(2012)]{langmead2012fast}
Ben Langmead and Steven~L Salzberg.
\newblock Fast gapped-read alignment with bowtie 2.
\newblock \emph{Nature methods}, 9\penalty0 (4):\penalty0 357, 2012.

\bibitem[Lauring et~al.(2013)Lauring, Frydman, and Andino]{lauring2013role}
Adam~S Lauring, Judith Frydman, and Raul Andino.
\newblock The role of mutational robustness in rna virus evolution.
\newblock \emph{Nature Reviews Microbiology}, 11\penalty0 (5):\penalty0 327,
  2013.

\bibitem[Li et~al.(2015)Li, Liu, Luo, Sadakane, and Lam]{li2015megahit}
Dinghua Li, Chi-Man Liu, Ruibang Luo, Kunihiko Sadakane, and Tak-Wah Lam.
\newblock Megahit: an ultra-fast single-node solution for large and complex
  metagenomics assembly via succinct de bruijn graph.
\newblock \emph{Bioinformatics}, 31\penalty0 (10):\penalty0 1674--1676, 2015.

\bibitem[Li and Durbin(2009)]{li2009fast}
Heng Li and Richard Durbin.
\newblock Fast and accurate short read alignment with burrows--wheeler
  transform.
\newblock \emph{Bioinformatics}, 25\penalty0 (14):\penalty0 1754--1760, 2009.

\bibitem[Li et~al.(2017)Li, Quang, and Xie]{li2017understanding}
Yi~Li, Daniel Quang, and Xiaohui Xie.
\newblock Understanding sequence conservation with deep learning.
\newblock In \emph{Proceedings of the 8th ACM International Conference on
  Bioinformatics, Computational Biology, and Health Informatics}, pages
  400--406. ACM, 2017.

\bibitem[Li et~al.(2016)Li, Shi, and Wasserman]{li2016genome}
Yifeng Li, Wenqiang Shi, and Wyeth~W Wasserman.
\newblock Genome-wide prediction of cis-regulatory regions using supervised
  deep learning methods.
\newblock \emph{bioRxiv}, page 041616, 2016.

\bibitem[Lu et~al.(2017)Lu, Chen, Fuhrman, and Sun]{lu2017cocacola}
Yang~Young Lu, Ting Chen, Jed~A Fuhrman, and Fengzhu Sun.
\newblock Cocacola: binning metagenomic contigs using sequence composition,
  read coverage, co-alignment and paired-end read linkage.
\newblock \emph{Bioinformatics}, 33\penalty0 (6):\penalty0 791--798, 2017.

\bibitem[Luo et~al.(2014)Luo, Rodriguez-r, and Konstantinidis]{luo2014mytaxa}
Chengwei Luo, Luis~M Rodriguez-r, and Konstantinos~T Konstantinidis.
\newblock Mytaxa: an advanced taxonomic classifier for genomic and metagenomic
  sequences.
\newblock \emph{Nucleic acids research}, 42\penalty0 (8):\penalty0 e73--e73,
  2014.

\bibitem[Ma et~al.(2018)Ma, You, Mai, Tokuyasu, and Liu]{ma2018human}
Yingfei Ma, Xiaoyan You, Guoqin Mai, Taku Tokuyasu, and Chenli Liu.
\newblock A human gut phage catalog correlates the gut phageome with type 2
  diabetes.
\newblock \emph{Microbiome}, 6\penalty0 (1):\penalty0 24, 2018.

\bibitem[Minot et~al.(2011)Minot, Sinha, Chen, Li, Keilbaugh, Wu, Lewis, and
  Bushman]{minot2011human}
Samuel Minot, Rohini Sinha, Jun Chen, Hongzhe Li, Sue~A Keilbaugh, Gary~D Wu,
  James~D Lewis, and Frederic~D Bushman.
\newblock The human gut virome: inter-individual variation and dynamic response
  to diet.
\newblock \emph{Genome research}, 21\penalty0 (10):\penalty0 1616--1625, 2011.

\bibitem[Nakatsu et~al.(2015)Nakatsu, Li, Zhou, Sheng, Wong, Wu, Ng, Tsoi,
  Dong, Zhang, et~al.]{nakatsu2015gut}
Geicho Nakatsu, Xiangchun Li, Haokui Zhou, Jianqiu Sheng, Sunny~Hei Wong,
  William Ka~Kai Wu, Siew~Chien Ng, Ho~Tsoi, Yujuan Dong, Ning Zhang, et~al.
\newblock Gut mucosal microbiome across stages of colorectal carcinogenesis.
\newblock \emph{Nature communications}, 6:\penalty0 8727, 2015.

\bibitem[Norman et~al.(2015)Norman, Handley, Baldridge, Droit, Liu, Keller,
  Kambal, Monaco, Zhao, Fleshner, et~al.]{norman2015disease}
Jason~M Norman, Scott~A Handley, Megan~T Baldridge, Lindsay Droit, Catherine~Y
  Liu, Brian~C Keller, Amal Kambal, Cynthia~L Monaco, Guoyan Zhao, Phillip
  Fleshner, et~al.
\newblock Disease-specific alterations in the enteric virome in inflammatory
  bowel disease.
\newblock \emph{Cell}, 160\penalty0 (3):\penalty0 447--460, 2015.

\bibitem[Paez-Espino et~al.(2016)Paez-Espino, Eloe-Fadrosh, Pavlopoulos,
  Thomas, Huntemann, Mikhailova, Rubin, Ivanova, and
  Kyrpides]{paez2016uncovering}
David Paez-Espino, Emiley~A Eloe-Fadrosh, Georgios~A Pavlopoulos, Alex~D
  Thomas, Marcel Huntemann, Natalia Mikhailova, Edward Rubin, Natalia~N
  Ivanova, and Nikos~C Kyrpides.
\newblock Uncovering earth’s virome.
\newblock \emph{Nature}, 536\penalty0 (7617):\penalty0 425, 2016.

\bibitem[Paez-Espino et~al.(2017)Paez-Espino, Pavlopoulos, Ivanova, and
  Kyrpides]{paez2017nontargeted}
David Paez-Espino, Georgios~A Pavlopoulos, Natalia~N Ivanova, and Nikos~C
  Kyrpides.
\newblock Nontargeted virus sequence discovery pipeline and virus clustering
  for metagenomic data.
\newblock \emph{Nature protocols}, 12\penalty0 (8):\penalty0 1673, 2017.

\bibitem[Peng et~al.(2010)Peng, Leung, Yiu, and Chin]{peng2010idba}
Yu~Peng, Henry~CM Leung, Siu-Ming Yiu, and Francis~YL Chin.
\newblock Idba--a practical iterative de bruijn graph de novo assembler.
\newblock In \emph{Annual international conference on research in computational
  molecular biology}, pages 426--440. Springer, 2010.

\bibitem[Peng et~al.(2012)Peng, Leung, Yiu, and Chin]{peng2012idba}
Yu~Peng, Henry~CM Leung, Siu-Ming Yiu, and Francis~YL Chin.
\newblock Idba-ud: a de novo assembler for single-cell and metagenomic
  sequencing data with highly uneven depth.
\newblock \emph{Bioinformatics}, 28\penalty0 (11):\penalty0 1420--1428, 2012.

\bibitem[Peterson et~al.(2009)Peterson, Garges, Giovanni, McInnes, Wang,
  Schloss, Bonazzi, McEwen, Wetterstrand, Deal, et~al.]{peterson2009nih}
Jane Peterson, Susan Garges, Maria Giovanni, Pamela McInnes, Lu~Wang, Jeffery~A
  Schloss, Vivien Bonazzi, Jean~E McEwen, Kris~A Wetterstrand, Carolyn Deal,
  et~al.
\newblock The nih human microbiome project.
\newblock \emph{Genome research}, 19\penalty0 (12):\penalty0 2317--2323, 2009.

\bibitem[Poplin et~al.(2017)Poplin, Newburger, Dijamco, Nguyen, Loy, Gross,
  McLean, and DePristo]{poplin2017creating}
Ryan Poplin, Dan Newburger, Jojo Dijamco, Nam Nguyen, Dion Loy, Sam~S Gross,
  Cory~Y McLean, and Mark~A DePristo.
\newblock Creating a universal snp and small indel variant caller with deep
  neural networks.
\newblock \emph{BioRxiv}, page 092890, 2017.

\bibitem[Quang and Xie(2016)]{quang2016danq}
Daniel Quang and Xiaohui Xie.
\newblock Danq: a hybrid convolutional and recurrent deep neural network for
  quantifying the function of dna sequences.
\newblock \emph{Nucleic acids research}, 44\penalty0 (11):\penalty0 e107--e107,
  2016.

\bibitem[Quang and Xie(2017)]{quang2017factornet}
Daniel Quang and Xiaohui Xie.
\newblock Factornet: a deep learning framework for predicting cell type
  specific transcription factor binding from nucleotide-resolution sequential
  data.
\newblock \emph{bioRxiv}, page 151274, 2017.

\bibitem[Rampelli et~al.(2016)Rampelli, Soverini, Turroni, Quercia, Biagi,
  Brigidi, and Candela]{rampelli2016viromescan}
Simone Rampelli, Matteo Soverini, Silvia Turroni, Sara Quercia, Elena Biagi,
  Patrizia Brigidi, and Marco Candela.
\newblock Viromescan: a new tool for metagenomic viral community profiling.
\newblock \emph{BMC genomics}, 17\penalty0 (1):\penalty0 165, 2016.

\bibitem[Ren et~al.(2017)Ren, Ahlgren, Lu, Fuhrman, and Sun]{ren2017virfinder}
Jie Ren, Nathan~A Ahlgren, Yang~Young Lu, Jed~A Fuhrman, and Fengzhu Sun.
\newblock Virfinder: a novel k-mer based tool for identifying viral sequences
  from assembled metagenomic data.
\newblock \emph{Microbiome}, 5\penalty0 (1):\penalty0 69, 2017.

\bibitem[Reyes et~al.(2015)Reyes, Blanton, Cao, Zhao, Manary, Trehan, Smith,
  Wang, Virgin, Rohwer, and {others}]{reyes2015gut}
Alejandro Reyes, Laura~V Blanton, Song Cao, Guoyan Zhao, Mark Manary, Indi
  Trehan, Michelle~I Smith, David Wang, Herbert~W Virgin, Forest Rohwer, and
  {others}.
\newblock {Gut DNA viromes of Malawian twins discordant for severe acute
  malnutrition}.
\newblock \emph{Proceedings of the National Academy of Sciences}, 112\penalty0
  (38):\penalty0 11941--11946, 2015.

\bibitem[Roux et~al.(2011)Roux, Faubladier, Mahul, Paulhe, Bernard, Debroas,
  and Enault]{roux2011metavir}
Simon Roux, Micha{\"e}l Faubladier, Antoine Mahul, Nils Paulhe, Aur{\'e}lien
  Bernard, Didier Debroas, and Fran{\c{c}}ois Enault.
\newblock Metavir: a web server dedicated to virome analysis.
\newblock \emph{Bioinformatics}, 27\penalty0 (21):\penalty0 3074--3075, 2011.

\bibitem[Roux et~al.(2015)Roux, Enault, Hurwitz, and
  Sullivan]{roux2015virsorter}
Simon Roux, Francois Enault, Bonnie~L Hurwitz, and Matthew~B Sullivan.
\newblock Virsorter: mining viral signal from microbial genomic data.
\newblock \emph{PeerJ}, 3:\penalty0 e985, 2015.

\bibitem[Roux et~al.(2016)Roux, Brum, Dutilh, Sunagawa, Duhaime, Loy, Poulos,
  Solonenko, Lara, Poulain, et~al.]{roux2016ecogenomics}
Simon Roux, Jennifer~R Brum, Bas~E Dutilh, Shinichi Sunagawa, Melissa~B
  Duhaime, Alexander Loy, Bonnie~T Poulos, Natalie Solonenko, Elena Lara, Julie
  Poulain, et~al.
\newblock Ecogenomics and potential biogeochemical impacts of globally abundant
  ocean viruses.
\newblock \emph{Nature}, 537\penalty0 (7622):\penalty0 689--693, 2016.

\bibitem[Singh et~al.(2016)Singh, Yang, Poczos, and Ma]{singh2016predicting}
Shashank Singh, Yang Yang, Barnabas Poczos, and Jian Ma.
\newblock Predicting enhancer-promoter interaction from genomic sequence with
  deep neural networks.
\newblock \emph{bioRxiv}, page 085241, 2016.

\bibitem[Storey et~al.(2003)]{storey2003positive}
John~D Storey et~al.
\newblock The positive false discovery rate: a bayesian interpretation and the
  q-value.
\newblock \emph{The annals of statistics}, 31\penalty0 (6):\penalty0
  2013--2035, 2003.

\bibitem[Truong et~al.(2015)Truong, Franzosa, Tickle, Scholz, Weingart,
  Pasolli, Tett, Huttenhower, and Segata]{truong2015metaphlan2}
Duy~Tin Truong, Eric~A Franzosa, Timothy~L Tickle, Matthias Scholz, George
  Weingart, Edoardo Pasolli, Adrian Tett, Curtis Huttenhower, and Nicola
  Segata.
\newblock Metaphlan2 for enhanced metagenomic taxonomic profiling.
\newblock \emph{Nature methods}, 12\penalty0 (10):\penalty0 902--903, 2015.

\bibitem[Vogtmann et~al.(2016)Vogtmann, Hua, Zeller, Sunagawa, Voigt, Hercog,
  Goedert, Shi, Bork, and Sinha]{vogtmann2016colorectal}
Emily Vogtmann, Xing Hua, Georg Zeller, Shinichi Sunagawa, Anita~Y Voigt, Rajna
  Hercog, James~J Goedert, Jianxin Shi, Peer Bork, and Rashmi Sinha.
\newblock Colorectal cancer and the human gut microbiome: reproducibility with
  whole-genome shotgun sequencing.
\newblock \emph{PloS one}, 11\penalty0 (5):\penalty0 e0155362, 2016.

\bibitem[Wang et~al.(2018)Wang, Tai, Wei, et~al.]{wang2018define}
Meng Wang, Cheng Tai, Liping Wei, et~al.
\newblock Define: deep convolutional neural networks accurately quantify
  intensities of transcription factor-dna binding and facilitate evaluation of
  functional non-coding variants.
\newblock \emph{Nucleic acids research}, 2018.

\bibitem[Wommack et~al.(2012)Wommack, Bhavsar, Polson, Chen, Dumas,
  Srinivasiah, Furman, Jamindar, and Nasko]{wommack2012virome}
K~Eric Wommack, Jaysheel Bhavsar, Shawn~W Polson, Jing Chen, Michael Dumas,
  Sharath Srinivasiah, Megan Furman, Sanchita Jamindar, and Daniel~J Nasko.
\newblock Virome: a standard operating procedure for analysis of viral
  metagenome sequences.
\newblock \emph{Standards in genomic sciences}, 6\penalty0 (3):\penalty0 421,
  2012.

\bibitem[Wood and Salzberg(2014)]{wood2014kraken}
Derrick~E Wood and Steven~L Salzberg.
\newblock Kraken: ultrafast metagenomic sequence classification using exact
  alignments.
\newblock \emph{Genome biology}, 15\penalty0 (3):\penalty0 R46, 2014.

\bibitem[Yue and Wang(2018)]{yue2018deep}
Tianwei Yue and Haohan Wang.
\newblock Deep learning for genomics: A concise overview.
\newblock \emph{arXiv preprint arXiv:1802.00810}, 2018.

\bibitem[Zeng and Gifford(2017)]{zeng2017predicting}
Haoyang Zeng and David~K Gifford.
\newblock Predicting the impact of non-coding variants on dna methylation.
\newblock \emph{Nucleic acids research}, 45\penalty0 (11):\penalty0 e99--e99,
  2017.

\bibitem[Zeng et~al.(2016)Zeng, Edwards, Liu, and
  Gifford]{zeng2016convolutional}
Haoyang Zeng, Matthew~D Edwards, Ge~Liu, and David~K Gifford.
\newblock Convolutional neural network architectures for predicting
  dna--protein binding.
\newblock \emph{Bioinformatics}, 32\penalty0 (12):\penalty0 i121--i127, 2016.

\bibitem[Zhou and Troyanskaya(2015)]{zhou2015predicting}
Jian Zhou and Olga~G Troyanskaya.
\newblock Predicting effects of noncoding variants with deep learning--based
  sequence model.
\newblock \emph{Nature methods}, 12\penalty0 (10):\penalty0 931, 2015.

\end{thebibliography}

\end{document}